\documentclass[prd, twocolumn, nofootinbib]{revtex4}
\usepackage{graphicx}
\usepackage{dcolumn}
\usepackage{epsfig} 
\usepackage{amsmath}
\usepackage{amsfonts}
\usepackage{graphicx}
\usepackage{amssymb}
\usepackage{color}

\begin{document}
\title{Testing the imprint of non-standard cosmologies on void profiles using Monte Carlo random walks}

\author{Ixandra Achitouv$^{1,2}$\footnote{E-mail:iachitouv@swin.edu.au}}

\address{ 
$^{1}$Centre for Astrophysics \& Supercomputing, Swinburne University of Technology, P.O. Box 218, Hawthorn, VIC 3122, Australia \\
$^{2}$ARC Centre of Excellence for All-sky Astrophysics (CAASTRO), 44 Rosehill St, Redfern, NSW 2016, Australia
}

\begin{abstract}

Using a Monte Carlo random walks of a log-normal distribution, we show how to qualitatively study void properties for non-standard cosmologies. We apply this method to an $f(R)$ modified gravity model and recover the N-body simulation results of \citep{ABPW} for the void profiles and their deviation from GR. This method can potentially be extended to study other properties of the large scale structures such as the abundance of voids or overdense environments. 
We also introduce a new way to identify voids in the cosmic web, using only a few measurements of the density fluctuations around random positions. This algorithm allows to select voids with specific profiles and radii. As a consequence, we can target classes of voids with higher differences between $f(R)$ and standard gravity void profiles. Finally we apply our void criteria to galaxy mock catalogues and discuss how the flexibility of our void finder can be used to reduce systematics errors when probing the growth rate in the galaxy-void correlation function.  

\end{abstract} 
\maketitle

\section{Introduction}
Over the past decade, galaxy surveys have revealed cosmic voids that are an essential component of the cosmic web (e.g.\ \citep{BondKofmanPogosyan96,Kirshneretal1981,Kauffmann&Fairall1991,Hoyle&Vogeley2002, Crotonetal2004, Panetal2012,Sutteretal2012,AB_voids2016}). Their dynamical formation carries information of the background expansion and the non-linear gravitational interactions, as matter flows out from underdense patches leading to the mass assembly of halos \citep{Dekel&Rees1994,Bernardeau&vandeWeygaert1997}. Therefore their statistical properties can be used to constrain cosmology, for instance using the integrated Sachs-Wolfe effect (e.g.\ \cite{Granettetal2009}), performing an Alcock-Paczynski test (e.g.\ \cite{Lavaux2012}), measuring their abundance or their density profiles (e.g.\ \cite{ANP,ABPW,Clampitt2013,Zivick2015}) or looking at the clustering of matter in underdense environments \citep{AB_BAO,Kirauta16}. 

\medskip
Furthermore, void statistics are promising to probe dark energy models such as coupled dark energy \cite{coupledDE} or modified gravity models that rely on screening mechanism, such as f(R) gravity \cite{HuSaw,chame}. In particular, the abundance of voids and void profiles can be used to test the cosmic expansion and the growth rate, where the fifth force is unscreened in underdense environments (e.g.\    \citep{HuSaw,ABPW,Cai2015,Li2012,Zivick2015}). However, precise theoretical predictions of how void abundance and void profiles change for modified gravity models is still lacking. This has driven intensive N-body numerical analyses of such properties (e.g.\ \cite{Zhaosim,Puchweinsim,Raserasim}). In this work, we introduce a fast estimate of how void profiles can vary for non-standard cosmologies, based on Monte Carlo Random Walks (MCRW). This method can potentially be explored to study other quantities such as the abundance of voids or the statistical properties of overdense peaks in the non-linear matter density field. 
\medskip

As an application, we show how different ways of selecting voids can enhance the imprint of $f(R)$ modified gravity, which has not been studied before. In fact, there is no one single approach to identify voids, nor should there be, as different techniques can highlight different properties of what each calls a void. For instance, many void finders define voids based on density criteria inside a sphere (e.g.\ \citep{Kauffmann&Fairall1991,Mulleretal2000,Hoyle&Vogeley2002, Colbergetal2005}). This definition is very helpful when trying to link the theory of an expanding underdense patch to the prediction of void abundance (for instance \cite{SVdW,ANP}). Other techniques based on watershed transforms or dynamical properties around voids (e.g.\ velocity field) have also been very useful to identify voids without imposing a particular shape for them (e.g.\ \citep{ElAd&Piran1997,Aikio&Mahonen1998,Plionis&Basilakos2002,Shandarinetal2006,AragonCalvoetal2007,Hahnetal2007, PlatenVdWJones2007}). With cosmological observations, ZOBOV/VIDE \citep{N08,VIDE} has also been quite successful in identifying density minima using Voronoi tessellation, leading to voids with interesting properties and that are not necessary spherical. In this case it is however more challenging to link the initial under-dense patches of matter to the identified void \cite{ANP}.

\medskip
In this work we study how a measurement of the density fluctuations at a limited number of scales is enough to identify voids, which would not be the case if the matter was not clustered. We will also show how the selection of voids with specific ridges can be important when probing the growth rate. 
  
\medskip
This paper is organized as follows: in Sec.~\ref{sec1}, we show how we can test the imprint on Large Scale Structure for non-standard cosmology using MCRW, focusing on void profiles. In Sec.~\ref{Appli1} we apply this technique to test the deviation of the void profiles for $f(R)$ gravity and discuss the importance of the void identification when testing for deviations with $\Lambda$CDM. In Sec.~\ref{sec2}, we apply the void finder criteria to mock catalogues, and show how a few measurements of the density fluctuations is enough to identify voids in a low density survey. Finally in Sec.~\ref{secAppli2}, we test the effect the identification criteria to the measurement of the growth rate, and highlight the advantages of having flexible void profiles (at a fixed void radius), when probing the growth rate. In Sec.~\ref{conclu} we present our conclusions.

\section{Theory}\label{sec1}
%We investigate how non-integrated 3+1 density criteria lead to voids using Monte-Carlo random walks.
In order to compute a local minimum in the dark matter density field, we generally need to measure the density contrast smoothed over some scale $\delta(R_S)$ and determine if it is below some density threshold. 

For instance, the spherical model (e.g.\ \citep{Gunn&Gott1973, SVdW}) of an underdense patch of matter expands linearly as the universe is expanding. If the initial patch is sufficiently deep to accumulate shells at the void boundary, the perturbation  becomes non-linear, its size increases faster than the background expansion, and shell crossing will occur leading to a void of a density contrast $\Delta(R_v)\sim -0.8$ for an Einstein de Sitter universe (where $R_v$ is the void radius). Hence some void finders naturally search for spherical patches that have a density contrast of $\sim -0.8$ once they are smoothed on a scale $R_S=R_v$. Such void finders require knowledge of the integrated density contrast over a scale $R_S$ and do not add any constraints to the void density profile. Void finders based on the watershed concept (e.g.\ \cite{N08,VIDE}) also need to have an estimate of the density around each particle (e.g.\ Voronoi Tessellation) to define zones that have a density minimum and thus voids. No constraints are imposed on the shape of the voids nor on their profiles. 

\medskip

In the next section we describe how we can use MCRW to study void profiles, by selecting a sample of trajectories that satisfy different density criteria. The choice of these criteria is not restrictive and can be tuned to target voids with specific characteristics. In addition, because our Universe is structured as the so called \textit{cosmic web}, we may wonder \textit{How much information do we need to have in order to find a void?}
For instance, considering a random position in a galaxy survey and measuring the density contrast at a smoothing scale $\delta(R_1)$ would it be enough to know if we are at the center of a void of size $R_v$? Note that in this case we do not consider the integrated density profile $\Delta(R_v)$ but rather the density profile at a given radial bin $\delta(R_1)$ which would be interesting to identify voids when looking at a galaxy survey with masked regions. The two quantities are linked by 

\begin{equation}
\Delta(R)=\frac{3}{R^3}\int_{0}^{R}\delta(q) q^2 dq. \label{intDel}
\end{equation}

The answer would naturally depend on the definition of the voids, thus let us consider voids as an underdense patch of matter (negative density contrast within the void) and a ridge that defines the void radius ($\delta(R_v)>0$). Let us also set a first sample of voids with radius $R_v\sim 20 \rm Mpc.h^{-1}$.

\subsection{Monte Carlo Random Walks}\label{secRDW}
%The log-normal distribution
%- introduce the concept of starting from log normal with condition
%show the RDW PDF for Sk

Considering a random position in a galaxy survey, the probability to find a density contrast $\Delta (R)$ smoothed on a scale $R$, is given by the the probability distribution function (PDF) of the cosmological density fluctuation. The full PDF carries all the non-linear gravitational interactions between the primordial density perturbations up to the present epoch. Hence it is the fundamental quantity that characterizes the clustering of matter in the Universe and all its hierarchical order (e.g.\ skewness). In the standard inflationary model (slow roll inflation with a single field), this PDF is initially Gaussian. Theoretical models that estimate the late time evolution of this PDF include perturbation theory (e.g.\ \citep{Bernardeau1994,Protogeros&Scherrer1997,Gaztanaga&Croft1999, Scherrer&Gaztanaga2001}) and the excursion set theory (e.g.\ \cite{Bondetal, Sheth1998, LamSheth2008}). Those methods provide a good physical understanding but are limited by the non-linear evolution as well as the mapping between the Lagrangian to Eulerian space that often assumes a deterministic spherical evolution \cite{Fosalba&Gaztanage1998}. From an empirical approach, the $1-$point PDF of galaxies is well described by a log-normal distribution (e.g.\ \citep{Hamilton1985, Bouchetetal1993, Kofmanetal1994}). This has been confirmed by several N-body simulations (e.g.\ \citep{Coles&Jones1991,Kofmanetal1994,Taylor&Watts2000, KayoTaruyaSuto2001}) even in the highly non-linear regime (down to $R\sim 2\rm Mpc.h^{-1}$ for $\Lambda$CDM cosmology \cite{KayoTaruyaSuto2001}). This is an impressive result considering that the dynamics of the initial density perturbations can only be described using N-body simulations. For this reason, in what follows we use the log-normal PDF to study the density criteria that identify voids with specific characteristics. Note that we will neglect the higher order of the full PDF, hence our results can only describe the qualitative feature of a full N-body simulation.

\medskip

\begin{figure}[ht]
\centering
\includegraphics[scale=0.4]{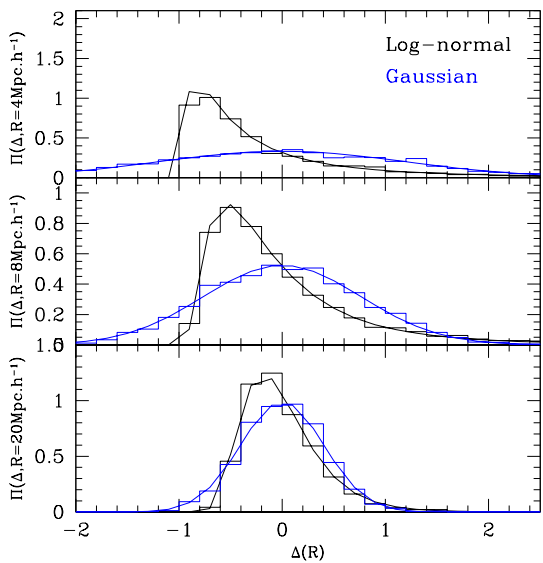} 
\caption{PDF of the Gaussian $\Delta(R)$ (blue histogram) and log-normal $\Delta_{\rm LN}(R)$ (black histogram) Monte Carlo Random walks at 3 different smoothing scales. The solid curve shows the associated Eq.(\ref{PDFG}, \ref{PDFLN}).}
\label{Fig1}
\end{figure}

For an initial matter density field with Gaussian statistics, the evolution of the density fluctuations as function of the smoothing scale follows a stochastic Langevin equation:

\begin{equation}
\frac{\partial \Delta (R,\textbf{x}=0)}{\partial R}=\int \frac{d^3 k}{2\pi^3}\;  \tilde{\delta}_k \; \frac{\partial \tilde{W}(k,R)}{\partial R} 
\end{equation}

where $\tilde{\delta}_k$ and $\tilde{W}(k,R)$ are the Fourier transforms of the density fluctuation and the filter function respectively. Instead of using the smoothing scale (or the linear variance) as an integration variable, \cite{Bondetal} have shown that it is more convenient to use $\ln{k}$ when solving for the evolution of the so called trajectory ($\Delta (R)$). Each trajectory is defined by a set of values for $\Delta={\Delta(R_0), \Delta(R_2),...\Delta(R_n)}$ and the stochastic Langevin equation to solve is

\begin{equation}
\frac{\partial \Delta (\textbf{x},R,\ln{k})}{\partial \ln{k}}=\eta(\textbf{x},\ln{k})\tilde{W}(k,R), \label{Langevin}
\end{equation}
where the stochastic force is a Gaussian white noise, 
\begin{equation}
\begin{split}
&\langle\eta(\textbf{x}_1,\ln{k_1}) \eta(\textbf{x}_2,\ln{k_2})\rangle=\\
&\delta_D(\ln{k_2}-\ln{k_1})P_{\rm Lin}(k_1) \frac{\sin{k_1 R}}{k_1 R}.
\end{split} \label{Langevin2}
\end{equation}

We can solve numerically the trajectories by performing a Monte Carlo approach. We use a large number of trajectories $\Delta(R,\ln{k})$ as function of the smoothing scale $R$. To construct each trajectory, we integrate the Langevin equation (Eq.\ref{Langevin}) over the logarithm wavenumbers $\ln{k}$, adding on each step the stochastic force \cite{Bondetal} This leads to the well known  Gaussian random walks \citep{Bondetal,ARSC,AWWR} that have a variance

 \begin{equation}
 \sigma^{2}_{\rm Lin}(R)=\frac{1}{2\pi^2}\int P_{\rm Lin}(k) \tilde{W}^2(k,R) k^2 dk\label{sigmalin}
 \end{equation}

For this analysis, we use the linear power spectrum from CAMB \footnote{http://camb.info/readme.html} \cite{CAMB} using a WMAP-5 cosmology \cite{wmap} ($\Omega_m=0.26, h=0.72, \sigma_8=0.79, n_s=0.963, \Omega_b=0.044$). This cosmology corresponds to the N-body simulations that we will use in Sec.~\ref{sec2}. 

In Fig.~\ref{Fig1} we can see the distribution of these random walks (Gaussian) at different smoothing scales (blue histogram). The solid blue PDF is the standard Gaussian PDF 

\begin{equation}
P(\Delta,\sigma_{\rm Lin})=\frac{1}{\sqrt{2\pi\sigma^{2}_{\rm Lin}(R)}}\exp\left( -\frac{\Delta^2}{2\sigma^{2}_{\rm Lin}(R)}\right). \label{PDFG}
\end{equation}

As we already mentioned, this PDF corresponds to the initial statistic of the matter density fluctuations. In order to obtain Monte Carlo walks that have a log-normal distribution, we simply use the mathematical correspondence between a Gaussian and a log-normal distribution  (e.g.\ \cite{KayoTaruyaSuto2001}), when we solve for Eq.~\ref{Langevin},\ref{Langevin2}:

\begin{equation}
\begin{split}
&\Delta_{\rm LN}+1=\frac{1}{\sqrt{1+\sigma_{\rm NL}^{2}(R)}}\times \\
&\exp\left(\frac{\Delta}{\sigma_{\rm Lin}(R)}\sqrt{\ln(1+\sigma_{\rm NL}^{2}(R))}\right),\label{mapping}
\end{split}
\end{equation}

where 

 \begin{equation}
 \sigma^{2}_{\rm NL}(R)=\frac{1}{2\pi^2}\int P_{\rm NL}(k) \tilde{W}^2(k,R) k^2 dk\label{sigmaNL}
 \end{equation}

This time, the subscript NL indicates that we use the non-linear power spectrum of the matter density field for which we use the halo-fit from CAMB \cite{CAMB} with the same WMAP-5 cosmology. In Fig~\ref{Fig1} we can see the corresponding Monte Carlo walks at 3 different smoothing scales (black histograms). The solid back curves show the corresponding log-normal distribution:

\begin{equation}
\begin{split}
&P(\Delta_{\rm LN},\sigma_{\rm NL}^{2}(R))=\frac{1}{\sqrt{2\pi\sigma_{\rm eff}^{2}}}\times\\
&\exp\left[-\frac{(\ln(1+\Delta_{\rm LN})+\sigma_{\rm eff}^{2}/2)^2}{2\sigma_{\rm eff}^2}\right]\frac{1}{1+\Delta_{\rm LN}},\label{PDFLN}
\end{split}
\end{equation}  
where $\sigma_{\rm eff}^{2}=\ln[1+\sigma_{\rm NL}^{2}(R)]$.

Unsurprisingly, we can see that on large scales (lower panel) the standard deviation of these PDFs is smaller than on small scales (top panel). This is a direct consequence of Eq.~\ref{sigmalin},\ref{sigmaNL}. In the limit where $R\rightarrow \infty$, $\sigma_{\rm NL}, \sigma_{\rm Lin}\rightarrow 0$ these PDFs become a Dirac delta functions centred on zero. This is satisfied by construction, as a consequence of the homogeneous universe on large scales while on small scales the matter density fluctuations (linear/non linear ones) can fluctuate significantly (e.g.\ if they correspond to a proto-halo/halo). Finally note that the choice of the filter does not matter when generating those random walks since we do not add any conditions such as an absorbing boundary threshold (used in the excursion set theory \cite{Bondetal}). However, the choice of filter has a physical meaning as it defines the smoothed volume that we consider ($V(R_S)=\int d^3 x W(x,R_S)$). Therefore, in what follows, we will only consider a top-hat filter in real space $\Theta(x-R_S)$, leading to $V(R_S)=\frac{4}{3}\pi R_S^3$.

\medskip
In the next section we will select a sample of those Monte Carlo walks that satisfied some density constrains relevant to identify voids.

\subsection{Density criteria to find voids}

We are interested in identifying voids at the present epoch hence, we want that at a random position in a galaxy survey, considering the matter within a small smoothing scale $R_m \rightarrow 0$, the corresponding density contrast $\Delta_{\rm LN}(R_m)$ tends to $-1$ (no matter). The choice of $R_m$ can be adjusted to give a smooth density profile $R_m \sim 1 \rm Mpc.h^{-1}$ or a  sharper profile: in the limit of a top-hat void profile, the matter density inside the void is null. Therefore the upper limit of  $R_m$ is the actual void size. In what follows we will consider the following example: $R_v=17.25\rm Mpc.h^{-1}$, $R_m=2\rm Mpc.h^{-1}$ and $\Delta_{\rm LN}(R<R_m)<-0.9$. We also require a ridge at the size of the void radius by adding the condition $\delta_{\rm LN}(R_v \pm \varepsilon)>0$ with $\varepsilon=1\rm Mpc.h^{-1}$. Note that for any trajectory 
$\Delta_{\rm LN}(R)$ (cumulative smoothed density profile on scale R), we obtain the equivalent trajectory $\delta_{\rm LN}(r)$ by differentiating $\Delta_{\rm LN}(R)$.

\medskip

Those two requirements define a $1+1$ condition: one that looks for an empty patch at a small smoothing scale and the other one that gives a lower limit on the amplitude of the density fluctuation at the ridge, corresponding to an overdense compensation wall at the void radius. This last constraint is not mandatory for certain void finders  (e.g.\ \cite{VIDE}), particularly for large voids, and therefore we distinguish this additional condition as a $+1$ condition.  

\medskip

Finally we require that $\Delta_{\rm LN}(R<R_v) <\Delta_{\rm LN}(R_v)$ in order to reduce the scatter around the averaged void profile ($2+1$ criteria). This means we compute $\Delta_{\rm LN}$ for every value of 
R in the range $0 < R < R_v$, and if any $\Delta_{\rm LN}$ does not satisfy that condition, we exclude the trajectory

\medskip

\begin{figure}[ht]
\centering
\includegraphics[scale=0.4]{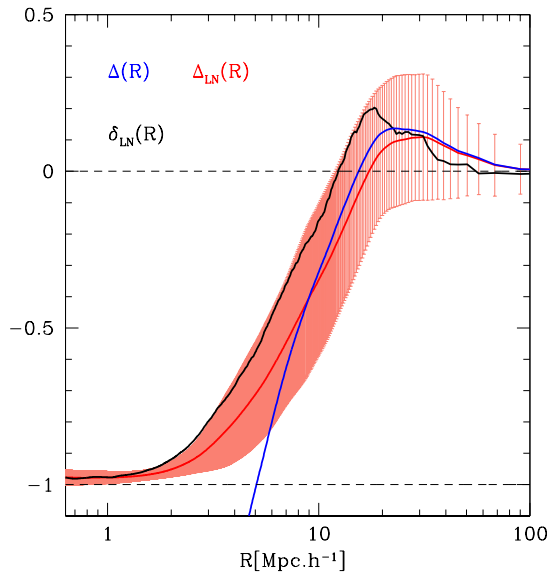} 
\caption{Averaged void profiles from Monte Carlo random walks for the non-linear integrated density fluctuations $\Delta_{\rm LN}(R)$ (red curve).  The light red bands show the standard deviation around this averaged profile. The corresponding linear density fluctuation is shown by the blue curve and the black curve shows the density fluctuation $\delta_{\rm LN}$ computed from the integrated. }
\label{Fig2}
\end{figure}

\begin{figure}[ht]
\centering
\includegraphics[scale=0.4]{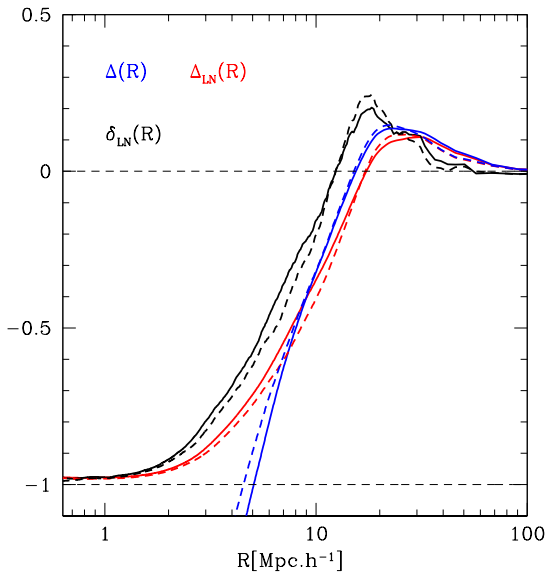} 
\caption{Averaged void profiles from Monte Carlo random walks for the non-linear integrated density fluctuations $\Delta_{\rm LN}(R)$ (red curve). The corresponding linear density fluctuation is shown by the blue curve and the black curve show the density fluctuation $\delta_{\rm LN}(R)$. The dotted curves show the same for $f_{R0}=-10^{-4}$. }
\label{Fig3}
\end{figure}

\begin{figure}[ht]
\centering
\includegraphics[scale=0.4]{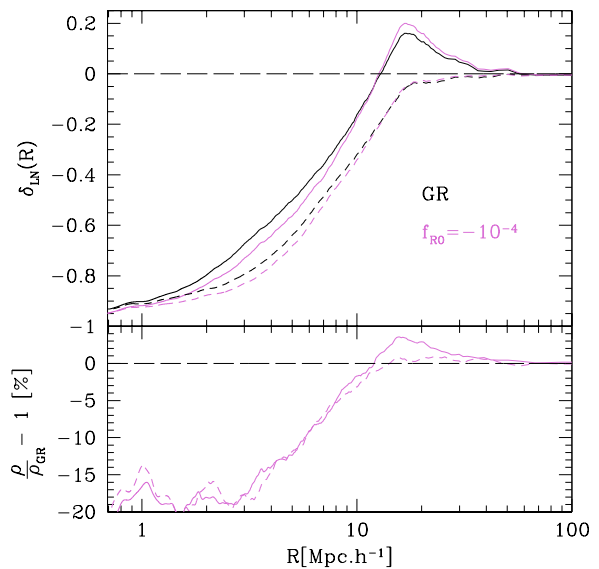} 
\caption{Averaged void profiles from Monte Carlo random walks for the non-linear density fluctuations $\delta_{\rm LN}(R)$ for GR (black curves) and $f_{R0}=-10^{-4}$ (violet curves) in the case of different density criteria to identify voids. The void profiles with a ridge are shown by the solid curves while the dashed curves show the ones without a ridge. The bottom panel show the ratio between the $f(R)$ density profiles with respect to the GR ones.}
\label{Fig4}
\end{figure}

Generating $1000$ trajectories, we found that $\sim 6.2\%$ of the trajectories satisfy these $2+1$ conditions. 

In Fig.~\ref{Fig2}, we can see the averaged value of all trajectories $\Delta_{\rm LN}$ (red curve) that satisfy these $2+1$ conditions while the light red band shows the standard deviation around the averaged mean value. The blue curve represents the mean value of the corresponding linear trajectories while the black curve shows the density fluctuation $\delta_{\rm LN}(R)=\frac{1}{3R^2}\frac{d(R^3\Delta_{\rm LN})}{dR} $.

This averaged fluctuation has what we can expect from a void profile at $z=0$: an underdensity at the center, slowly reaching a maximum density contrast on the ridge and having $\delta_{\rm LN}(R\gg Rv)\rightarrow 0$ on large scales. In principle, we could add more conditions to further reduce the scatter around the mean value of the void density profile. For instance we could add an upper limit on the amplitude of the ridge. However, if one wants to keep the criteria to a minimum, these $2+1$ criteria are a good compromise to obtain a mean density profile consistent with a void profile expectation.

\medskip

It is interesting to mention that the excursion set theory uses the linear density fluctuation to predict the abundance of voids (e.g.\ \citep{Bondetal,SVdW,ANP}). 
The spherical evolution (e.g.\ \cite{Gunn&Gott1973, SVdW}) of a linearly extrapolated under-density that corresponds to a void today is given by $\Delta (R) \sim -2.7$. This value is often used in the excursion set theory (e.g.\ \citep{Bondetal,SVdW,ANP}) to predict the abundance of voids, (when a random linear trajectory crosses that threshold at the largest smoothing scale without crossing the linear threshold of halo formation). One may be tempted to establish a link between the log-normal random walks that are identified as voids and the corresponding linear trajectories. However there are a few caveats. First, any deterministic link between the initial conditions and the non-linear density fluctuations is in theory not correct. While the non-linear density fluctuations can effectively be described by a log-normal distribution, there is no physical reason why it should be the case. Hence, using a log-normal transformation from initial Gaussian density fluctuations to study the link between the void identified today and the initial density criteria is in principle not a physical description of the non-linear processes leading to the formation of voids. Keeping this in mind, we may still investigate if there is an effective link between the two. However, the spherical criterion is a deterministic prediction that can not encapsulate all of the non-linear interactions of the density fluctuations. For instance in \cite{ANP}, the authors have measured, from an N-body simulation, the extrapolated linear density contrasts that lead to the voids identified with ZOBOV \cite{N08}. They show that on average and for large voids, the critical density was consistent with the spherical prediction of $\Delta \sim -2.7$ but showing a non-negligible scatter around this mean value. In addition, in the context of the excursion set theory, the averaging of trajectories at random positions requires an effectively lower density criteria than what is expected from density peak fluctuations \citep{Robertson2008,ARSC, AWWR} (by a factor $\sim 1/1.4$). 

\medskip
Hence, it would be interesting to investigate the consequences for the prediction of the abundance of voids, thus building an effective mapping between the void size we identify previously and the scale where the linear random walk trajectory crosses some effective threshold (e.g.\ $\Delta \sim -2.7/1.4$). This however,  goes beyond the scope of this work. 

\section{Application: void profiles departure from GR in $f(R)$ gravity using MCRW }\label{Appli1}
Modification of general relativity (GR) on large scales has been investigated as an alternative to the cosmological constant (e.g.\ \citep{HuSaw,chame}.  The f(R) gravity model is one example where a function of the Ricci scalar, $f(R)$ is added to the Einstein-Hilbert Action. The function $f(R)$ can be tuned to have the same expansion history as the standard $\Lambda$CDM scenario \citep{HuSaw}. Furthermore, to ensure the validity of GR in our local environment, the so-called Chameleon screening mechanism \cite{Khoury2004} is required. The former suppresses the deviation from GR in high-density environments such as the Milky Way. 

\medskip

In such models (e.g.~\cite{HuSaw}), the function $f(R)$ is given by
\begin{equation}
f(R)=-m^2 \frac{c_1 (R/m^2)^n}{c_2 (R/m^2)^n+1}
\end{equation}
where $m^2=H_0^2 \Omega_m$ with $H_0$ and $\Omega_m$ the Hubble parameter and matter density at $z=0$. The parameters $n,c_1$ and $c_2$ are free. To recover the background expansion close to $\Lambda$CDM, $f(R)$ is expressed as $f(R)=-6 m^2 \Omega_\Lambda/\Omega_m +O ((m^2/R)^n)$ and in what follows we use a $n=1$ model. In this case $f_R\equiv \frac{df(R)}{dR}$ can be interpreted as a scalar degree of freedom. A field equation can be obtain for $f_R$ where the only degree of freedom is set by the background field amplitude at $z=0$, $f_{R0}$. 

\medskip
The value of $\mid{f_{R0}}\mid$ controls the screening mechanism: smaller values of $\mid{f_{R0}}\mid$ correspond to higher screening. 
Current constraints from large scales and galaxy clusters rule out models with $\mid f_{R0}\mid\geq 10^{-4}$ \citep{Lombriser14,Terukina14,Jain13}. Nevertheless, In what follows we consider $f_{R0}= -10^{-4}$ for our analysis to test how our MCRW approach can reproduce void profiles in the case of modified gravity.

\medskip

In order to generate our MCRW for the $f(R)$ modified gravity, we use the halo fit from MGcamb \cite{MGcamb} with the cosmological parameters defined in Sec.~\ref{sec1} to compute Eq.(\ref{sigmaNL}), and let the other density criteria unchanged. Over $1000$ trajectories, $9\%$ satisfied the density criteria, an increase of $3/2$ compared to the number of trajectories for the GR case. This is in qualitative agreement with the result of N-body simulation of \citep{ABPW,Li2012,Cai2015,Zivick2015} where the author found more large voids for $f(R)$ gravity due to the fifth force that is unscreened in underdense environment. Physically, the fifth force pushes the particles stronger toward the walls of the voids \cite{Clampitt2012}. This means that cosmic voids are more efficient to form and that the $f(R)$ voids have a higher ridge amplitude. This has also been confirmed by \cite{ABPW,Li2012,Cai2015,Zivick2015}. 

\medskip

In Fig.~(\ref{Fig3}), we can see the comparison between GR density profiles (solid curves) discussed previously and the ones computed for $f(R)$ gravity with $f_{R0}=-10^{-4}$ (dotted curves). This shows the characteristic feature of $f(R)$ voids: the dotted lines are bellow the GR ones within the voids while on the ridge these density fluctuations become higher. It is quite remarkable that a MCRW approach can reproduce this main feature observed in an N-body simulation (e.g.\ \citep{ABPW,Li2012,Cai2015,Zivick2015}.). {One might wonder if the differences between the GR and $f(R)$ profiles are significant given the uncertainties in the profile shown in Fig.\ref{Fig2}. In Fig\ref{Fig2}, the scatter band is due to cosmic variance, (it is not a statistical error). When we generate the GR and $f(R)$ profiles, we use the same initial conditions (random seed and linear power spectrum) such that the cosmic variance cancelled out and the difference between the dashed Vs solid curves in Fig.\ref{Fig3}, \ref{Fig4} are directly due to the $f(R)$ imprint induced by a different non-linear power spectrum.\color{black}{}
Finally, the difference between the linear integrated density fluctuations (blue curves) is also interesting. They correspond to the Gaussian perturbation that lead to the identified voids and are different even though we started from the same initial conditions. The difference between the two is due to the additional voids identified in the $f(R)$ MCRW, leading to a different linear density threshold for the GR and $f(R)$ voids. This could be investigated further to predict the abundance of voids in non-standard cosmologies. 

\medskip

Additionally we can investigate the effect of the void identification criteria on these profiles. In fact previous studies of $f(R)$ imprints on void profiles have not investigated the effect of the void finder itself. As we already mentioned void finders such as \cite{N08,VIDE} tend to identify voids without ridge for $R_v\geq 17 \rm Mpc.h^{-1}$ (see \cite{Hammaus_vprofile}). Because the fifth force enhances the void ridge, it might be interesting to test the imprint of $f(R)$ for different types of void profiles (for a fixed $R_v$). In Fig.~\ref{Fig4} we can see the non-linear density fluctuation for GR (black curves) and $f(R)$ (violet curves) for two different voids that satisfy different criteria: voids that have a ridge (solid curves) and voids without (dashed lines).
For this example, to select voids with a ridge, we require that $\delta_{LN}(R_v=17.25\pm 1\rm Mpc.h^{-1})<-0.1$, $\Delta_{LN}(R<16.25\rm Mpc.h{-1})<\Delta_{LN}(R_v)$, and $\Delta_{LN}(R<2\rm Mpc.h^{-1})<-0.5$. The ridge condition is weaker than the previous condition in order to enhance the difference between GR and f(R) voids. For the same reason, the condition at low radius is less restrictive ($<-0.5$) compared to the previous example. For the voids without a ridge the conditions are $\delta_{LN}(R_v=17.25\pm 1\rm Mpc.h^{-1})<-0.4$, $\Delta_{LN}(R<16.25\rm Mpc.h{-1})<\Delta_{LN}(R_v)$, and $\Delta_{LN}(R<5\rm Mpc.h^{-1})<-0.5$. Both show the expected feature of the $f(R)$ voids: the inner part of the voids is again steeper than in GR.  

\medskip
In the lower panel we can see the relative difference between the $f_{R0}=-10^{-4}$ profiles and the corresponding GR profiles. The $f(R)$ voids with a ridge (solid line) show a departure from GR at the ridge while the ones without, only differ from GR in the inner part of the void profiles. Another way to explain this trend is because the clustering of the matter is more effective for $f(R)$ (e.g.\ \cite{ABPW}), voids will be more empty in the inner part (mass conservation) and will accrete more matter at their ridge. This is why the MCRW can be used to test the departure from GR. 

\medskip
This work indicates that requiring voids with a ridge might be important when probing non-standard gravity models. Our MCRW study can also be applied to study other statistical properties: we can study larger under-dense or overdense regions by selecting random walks of a given density fluctuation on large scales. In the next section we test further the advantages of having a flexible void finder, applied to galaxy mock catalogues.

\section{Void finder for galaxy mock catalogue}\label{sec2}

In this section, we use the  freely available DEUS N-body simulations used for several purposes (e.g.\ \citep{Alimi2010,Courtin2011,Rasera2010,ARSC,AWWR}). This simulation has a $648 \rm Mpc.h^{-1}$ box size with $2048^3$ particles, and was realized using the RAMSES code \cite{Teyssier2002} for a $\Lambda$CDM model calibrated to WMAP 5-yr parameters $(\Omega_m,\sigma_8,n_s,h)=(0.26,0.79,0.96,0.7)$. We built 36 dark matter and 36 galaxy mocks catalogues by sub-sampling $N_h=15000$ dark matter particles/halos. The halos are identified with the Friend-of-Friends (FoF) algorithm with linking length $b=0.2$, selecting the most massive haloes and leading to a mean density of $\bar{n}=0.003 \rm Mpc^{-3}.h^{3}$. These choices approximately mimic the selection of the 6dFGS galaxy survey \footnote{http://www.6dfgs.net/} and correspond to the choices made in \cite{AB_voids2016}. For the Dark Matter mocks, we randomly select a sample of dark matter particles in each catalogue until the density equals $\bar{n}=0.003 \rm Mpc^{-3}.h^{3}$.

\subsection{Method}\label{secMethod}

Previously we have used different density criteria to identify voids, using conditions both on the integrated density contrast $\Delta_{\rm LN}$ and the density fluctuation $\delta_{\rm LN}$. For mock catalogues, it becomes interesting to mainly probe the density fluctuation $\delta_{\rm LN}(R)$ at different scales in order to identify voids. Indeed, in such a case we do not have to assume any volume (shape for the voids). Furthermore, in a galaxy survey, some regions might be masked. In such case, computing the averaged profile $\delta_{\rm LN}(R)$ is generally done by counting pairs (using the correlation function that give the excess probability of having galaxies distant from $R+dR$ with respect to the mean density $\bar{\rho}$). Hence $\delta_{\rm LN} (R)\equiv \frac{\rho_{vg}(R)}{\bar{\rho}}-1=\xi_{vg}(R)$ (e.g.\ sec2.2 in\cite{Hammaus2015JCAP}).
\medskip

Given the position of dark matter particles or a number $N_{\rm Gal}$ of galaxies with coordinates $\mathbf{X}^j_{\rm G}=(x^j_{\rm G}, y^j_{\rm G},z^j_{\rm G})$, where $j=[1,N_{\rm Gal}]$. The previous density criteria to identify voids can be applied in the following steps:
\medskip

(i) Generate a number $N_{\rm Ran}$ of random positions that follow the selection function of the 
galaxies positions: $\mathbf{X}^l_{R}=(x^l_{\rm R}, y^l_{\rm R}, z^l_{\rm R})$, in order to determine the correlation function. Generate another uniform random set $N_{\rm Trial}$ of positions that span the spatial 
coverage of the galaxy catalogue, but do not have to trace the selection 
function $\mathbf{X}^j_{\rm T}=(x^j_{\rm T},y^j_{\rm T},z^j_{\rm T})$.
 \medskip
 
(i*) For an N-body simulation, $\mathbf{X}^l_{\rm R}$  and $\mathbf{X}^j_{\rm T}$ follow the same random distribution with min/max coordinates given by the size of the box. In the case where we only know the galaxy positions and the survey has masked regions, we can use the random distribution of the galaxies (if provided), to generate one for $\mathbf{X}^j_{\rm T}$. The idea is to count how many randoms are in a cell of length $dL<R_v$. Then we label all the cells that have a density above the average number per cell (above $1\sigma$ for instance).  Finally we draw a random distribution for $\mathbf{X}^j_{\rm T}$ and select only the positions that correspond to a labelled cell until we reach a total number of randoms equal to $N_{\rm Trial}$. Hence the position $\mathbf{X}^j_{\rm T}$ will avoid being next or in the masked regions. 
 
 \medskip
 
(ii). Compute for each positions $\mathbf{X}^j_{\rm T}$:\\
- the number of {Random Trial}\color{black}{} pairs $\rm RT (j,R_i)$ in bin
\begin{equation}
R_i=\sqrt{(x_{R}-x_T^j)^2+(y_{R}-y_T^j)^2+(z_{R}-z_T^j)^2}. \nonumber
\end{equation}

- the number of {Data (galaxy) Trial}\color{black}{} pairs $\rm DT (j, R_i)$ in bin 
\begin{equation}
R_i=\sqrt{(x_{\rm G}-x^j_{\rm T})^2+(y_{\rm G}-y^j_{\rm T})^2+(z_{\rm G}-z^j_{\rm T})^2}, \nonumber
\end{equation}
where $R_i \leq R_v$.

\medskip

(iii) Do two nested loops over the coordinate $j=[1,N_{\rm Trial}]$ and the separation pairs $R_i=[0,R_v]$. Within these loops, flag the coordinates $\mathbf{X}_{\rm T}^j$ that satisfied the criteria on the ratio:
\begin{equation}
\delta^j_{\rm T,G}=\frac{\rm DT (j, R_i)}{\rm RT (j,R_i)}\frac{N_{\rm Ran}}{N_{\rm Gal}}-1.\label{deltabin}
\end{equation}
The flagged coordinates are the void positions

\medskip

(iii*) Optional: Do an additional loop over the voids to exclude overlapping ones (trial positions which are closer together than $2R_v$) 

\medskip

\begin{figure}[ht]
\centering
\includegraphics[scale=0.4]{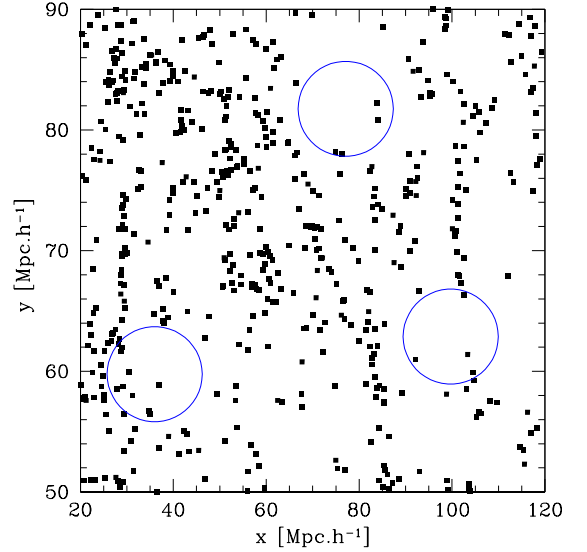} 
\caption{Sample of the voids with the $3+1$ criteria (blue circles) for a slice of a mock $dz=20\rm Mpc.h^{-1}$. Galaxies are shown by the black squares. }
\label{Fig5}
\end{figure}

In Fig.~ \ref{Fig5} we can see the result of the previous steps in a 
small sub-section of one of the galaxy mock catalogues, setting $R_v=17.5\rm Mpc.h^{-1}$ and using the density criteria used in \cite{AB_voids2016}: 
\begin{eqnarray}
\delta^j_{\rm T,G}(j,R=1\pm 1 \rm Mpc.h^{-1})<-0.9\label{c1} \\
\delta^j_{\rm T,G}(j,R=2\pm 1\rm Mpc.h^{-1})<-0.7\label{c2}, \\
\delta^j_{\rm T,G}(j,R=R_v+\Delta R )>\delta^j_{\rm T,G}(j,R_v)\label{c3} \;\rm{and}\\
 \delta^j_{\rm T,G}(j,R=R_v\pm 1 \rm Mpc.h^{-1})\geq 0 \label{cp1} \\
\end{eqnarray}

The first two conditions are similar to $\Delta_{\rm LN}(R<2\rm Mpc.h^{1})<-0.9$ for low density sample. The third and fourth (Eq.\ref{c3}\ref{cp1}) conditions ensure a ridge for the voids. In what follows we will refer to these conditions as the $3+1$ conditions. The choice of $N_{\rm Trial}$ is arbitrary: the higher $N_{\rm Trial}$, the higher number of voids is expected until it converges if we require non-overlapping voids (step iii*). In Fig.~\ref{Fig5}, we required non-overlapping voids and we find $N_{\rm Void}\sim 90$ voids using $N_{\rm Trial}\sim 10 N_{\rm Gal}$, $N_{\rm Ran}\sim 10 N_{\rm Gal}$ (case a). Neglecting (iii*), choosing $N_{\rm Trial}\sim 10 N_{\rm Gal}$, $N_{\rm Ran}\sim 10 N_{\rm Gal}$ leads to $N_{\rm Void}\sim 1500$ voids (case b). 

\medskip

Using $10$ mocks we employ the Landy-Szalay estimator \cite{LSestimate} to compute the LS cross-correlation function:
\begin{equation}
\begin{split}
\xi_{vg}(R)&=\frac{D_{\rm v}D_{\rm g}}{R_{\rm v}R_{\rm g}}\frac{N_{\rm Ran}N_{\rm Ran2}}{N_{\rm Gal} N_{\rm Void}} -\\
&\frac{D_{\rm v}R_{\rm v}}{R_{\rm v} R_{\rm g}}\frac{N_{\rm Ran}}{N_{\rm Void}}-\frac{D_{\rm g}R_{\rm g}}{R_{\rm v}R_{\rm g}}\frac{N_{\rm Ran2}}{N_{\rm Void}}+1. 
\end{split}
\end{equation}
where $N_{\rm Void}$ corresponds to the number of voids identified with the required density criteria and $N_{\rm Ran2}$ correspond to a random set of values that overlap with the voids. {The number of pairs at a distance $R$ are labelled by $\rm D_g, D_v$ for the galaxy and void data respectively while  $\rm R_g, R_v$ correspond galaxy and void pairs computed from the random distributions. }\color{black}{} 

\begin{figure}[ht]
\centering
\includegraphics[scale=0.4]{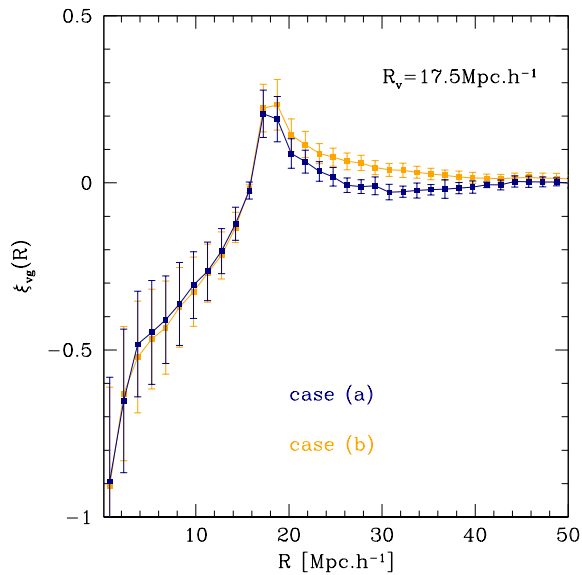} 
\caption{Mean density profile for the voids identified with the $3+1$ criteria using 10 mocks. }
\label{Fig6}
\end{figure}

In Fig.~\ref{Fig6} we can see the LS cross-correlation function for the identified voids, selecting non-overlapping voids (case a) and all voids (case b). The error bars are computed using the standard deviation over the mocks. The main effect of selecting non-overlapping voids is to reduce the amplitude of the correlation on scales $R>R_v$. Interestingly, the standard deviation of these mean density profiles is similar even though there are $\sim 15$ times more overlapping voids, demonstrating that overlapping voids do not add more information.

\subsection{Extension to different ridges and void sizes}
\begin{figure}[ht]
\centering
\includegraphics[scale=0.4]{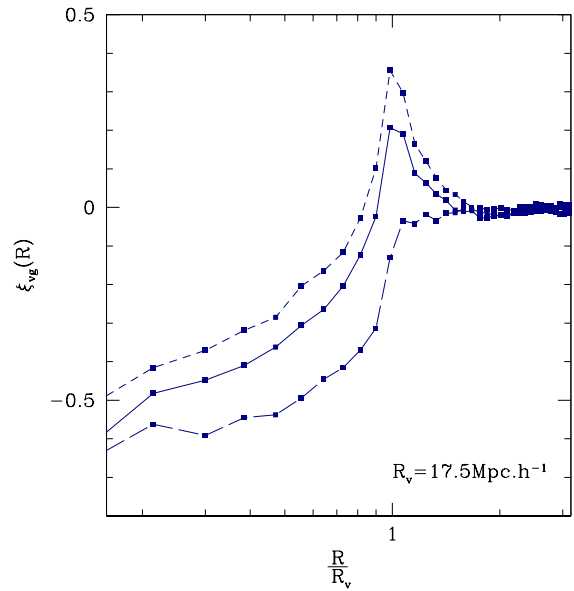} 
\caption{Mean density profile for the voids identified with the $3+1$ criteria using 10 mocks. The solid curve corresponds the $+1$ condition $\delta(R=R_v)>0$ (same as Fig.~\ref{Fig6}), the long-dashed curve corresponds to $\delta(R=R_v)>-0.5$ while the short-dashed curve corresponds to $\delta(R=R_v)>0.2$.}
\label{Fig_amp}
\end{figure}

\begin{figure}[ht]
\centering
\includegraphics[scale=0.4]{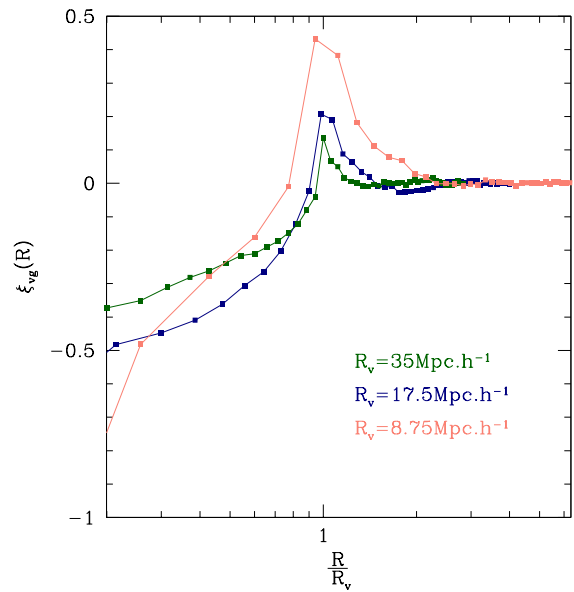} 
\caption{Mean density profile for the voids identified with the $3+1$ criteria defined in Sec.~\ref{secMethod} and for different void sizes. }
\label{Fig_size}
\end{figure}

One of the criteria we may want to vary is the amplitude of the void ridges. Previously we required $\delta(R=R_v)>0$, however we can require a higher ridge or no ridge at all. In Fig.~\ref{Fig_amp} we can see how the $+1$ condition changes once we choose $\delta(R=R_v)>-0.5$ (long dashed curve), $\delta(R=R_v)>0.2$ (short dashed curve) and $\delta(R=R_v)>0$ (solid curve) for the non-overlapping voids (case a). 

\medskip

More interestingly, we can also vary the void sizes $R_v$, keeping the other criteria. Given the hierarchical clustering of galaxies, we may wonder if our $3+1$ criteria, applied on the first 2 bins and around the ridge, are enough to identify large voids. Indeed, considering voids as twice the size of the previous ones for instance $R_v\sim 35 \rm Mpc.h^{-1}$, \textit{is probing the density on scales $R<2Mpc.h{-1}$ and on $R\sim 35\rm Mpc.h^{-1}$ enough to identify voids?} The answer to this question is positive as we can see from Fig.~(\ref{Fig_size}), where the green curve corresponds to voids with $R_v=35\pm 1 \rm Mpc.h^{-1}$. This would not be the case if the galaxies were randomly clustered, as we will show in the next section. The voids in Fig.~(\ref{Fig_size}) have the same $3+1$ criteria and are non-overlapping (case a), only $R_v$ varies (blue, pink and green curves correspond to $R_v=17.5\pm 1 \rm Mpc.h^{-1}$, $35\pm 1 \rm Mpc.h^{-1}$ and $8.75\pm 1 \rm Mpc.h^{-1}$ respectively). 
We note that the ridge is higher for smaller voids, in agreement with the expectation that galaxies are more clustered on small scales. However one could vary the $+1$ criterion to select specific void ridges. If we want a steeper profile for the large voids, we could change the second condition $\delta(R=2\pm 1)<-0.7$ to $\delta(R=5\pm 1)<-0.7$ while if we want a higher amplitude on the ridge, we can require $\delta(R_v)>0.2$, for instance.

\subsection{Discussion and limitation of the method}
\begin{figure}[ht]
\centering
\includegraphics[scale=0.4]{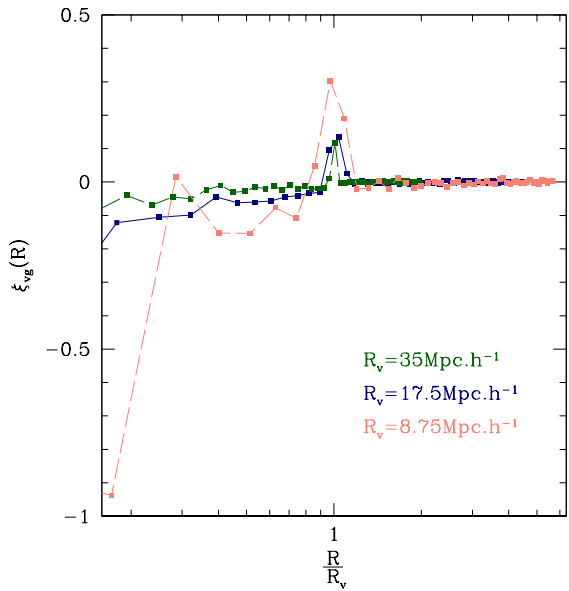} 
\caption{Mean density profile for the spurious voids identified with the $3+1$ criteria defined in Sec.~\ref{secMethod} and for different void sizes. }
\label{Fig_spurious}
\end{figure}

In this section, we have presented an easy way to identify voids in mock catalogues and in galaxy survey (see \cite{AB_voids2016} for an application of this void finder to the 6dFGS data). The method relies on probing $3+1$ density measurements in Eq.(\ref{deltabin}) at small $(R\leq 2\rm Mpc.h^{-1})$ scales and at the void scale $(R=R_v \rm Mpc.h^{-1})$. Furthermore, those criteria can be changed according to the choice of void profiles we target. This is a consequence of the cosmic clustering and using a low density mock catalogue. 
In fact, probing only $3+1$ bins in order to obtain smooth void profiles (as shown in Fig.~\ref{Fig_amp}, Fig.~\ref{Fig_size}), would not work if we consider a random set of positions. Indeed, if we replace the positions of the mock galaxies by random positions with the same number density, we will still find positions that satisfy the density criteria at the $3+1$ bins we consider. Those spurious voids would be due to Poisson fluctuations and would have a density profile as shown in Fig.~\ref{Fig_spurious} for different $R_v$ and using the same  $3+1$ criteria. As we can see, the selected spurious voids satisfied the required criteria but do not have a \textit{void profile} with a smooth shape as presented in Fig.~\ref{Fig_size}. {These spurious voids are also present in other void finder algorithms, such as ZOBOV. Using a higher density sample or choosing a more restrictive conditions to identify voids, such as lowering the density at the void centre or requiring conditions over the cumulative density fluctuation $\Delta_{\rm LN}$, would reduce their fractions.}\color{black}{} 

\section{Application: Systematic errors in the measurement of the growth rate for voids with different ridges}\label{secAppli2}

\begin{figure}[ht]
\centering
\includegraphics[scale=0.4]{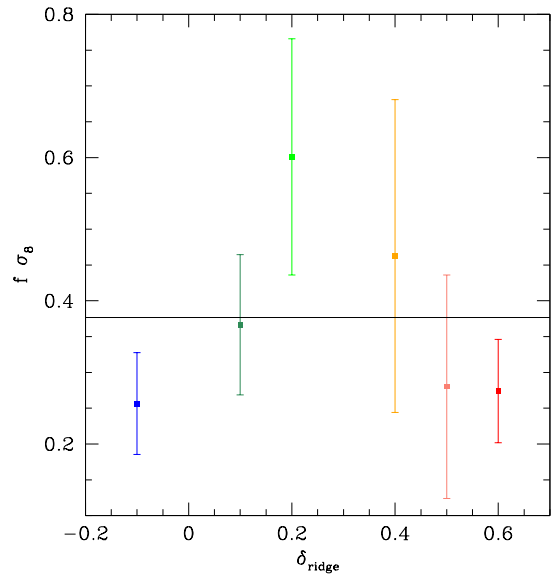} 
\caption{Best fitting values for $f\sigma_8$ as function of the ridge amplitude. The different colors correspond to voids with the same radius but a density fluctuation at the ridge $\delta_{\rm ridge}=-0.1,0.1,0.2,0.4,0.5,0.6$ for the blue, dark green, green, orange, pink and red squares respectively. }\label{figBF_f}

\end{figure}

Redshift space distortions around cosmic voids can be used to probe the growth rate $f\equiv \frac{d\ln{\delta_m(a)}}{d\ln{a}}$, with $\delta_m(a)$ the growing mode of matter density fluctuations and $a$ the scale factor. The growth rate is a powerful cosmological probe that is sensitive both to the cosmic expansion and the gravitational interactions between galaxies. On large scales, the linear growth rate can be measured by probing the coherent infall or outflow of galaxies sourced by the gravitational potential.

\medskip

 In the Gaussian streaming models (GSM) (e.g.\ \citep{Reid2011,KoppUA}) the void-matter correlation function in the local universe can be expressed as
 \begin{equation}
 \begin{split}
& \xi_{\rm v-DM}=\int (1+\xi_{\rm v-DM}^{1D}(y)) \times\\
& P\left(v-v_p(y) \left[\pi-\frac{v/H_0}{y}\right]\right)dv -1,
\end{split}\label{GSMeq}
 \end{equation}
where $v_p$ is the peculiar velocity of dark matter, $\pi$ is the line of sight, $y=\sqrt{\sigma^2+(\pi-v/H_0)^2)}$, with $\sigma$ the perpendicular direction to the line of sight and $P(v) dv$ is the stochastic velocity distribution of matter within a group (reproducing the small scale elongation along the line of sight). In the linear approximation, the peculiar velocity can be expressed as \cite{Peebles} $v_p(r)=-1/3 H_0 r\Delta_{\rm NL}(r) f$. The velocity distribution is given by 

\begin{equation}
P(v)dv=\frac{1}{\sqrt{2\pi\sigma_v}} \exp\left[-\frac{v^2}{2\sigma_v}\right] dv,
\end{equation}
where $\sigma_v(R)$ is the velocity dispersion. In what follows we neglect the scale dependence of $\sigma_v$ \citep{Hammaus2015JCAP,KoppUA,AB_voids2016}.

\medskip

To test the effect of the ridge amplitude without introducing complication due to the linear bias approximation (when probing the galaxy redshift space distortion), we consider only the measurement of the growth rate using the Dark Matter density mocks we introduced in section \ref{sec2}. 
We start by identifying voids in the Dark matter mock catalogues using the criteria from Eq. \ref{c1}, \ref{c2}, \ref{c3}, \ref{cp1}. Then we measure the mean density profile and use Eq.{\ref{intDel}} to obtain the integrated density profile in real space that we use in the GSM. Finally, we use the same dark matter density mocks to build the redshift space (RS) mock catalogues. To do so we shift the real space dark matter particles $\mathbf{r}$ to $\mathbf{s}$ using the flat sky approximation:

\begin{equation}
\mathbf{s}=\mathbf{r}+\frac{v_p(\mathbf{r})}{H_0}\mathbf{u}_r
\end{equation}
where $\mathbf{u}_r$ is the unitary vector along the line the sight (here the $z$-coordinate) and $v_p\equiv\mathbf{v}.\mathbf{u}_r$ is the peculiar velocity of the particles along the $z$-direction.  

\medskip

We identify voids in the RS catalogues and measure the averaged correlation function over 6 mocks $\sigma_6$, in redshift space $\left<\xi_{v-DM}(\pi,\sigma)\right>_6$ such that we have $36/6=6$ measurements of the averaged correlation function and the standard deviation averaged over the 6 mocks to compute the likelihood between the mock means and the GSM, letting $(f\sigma_8,\sigma_v)$ as free parameters in Eq.\ref{GSMeq}. To find the best fitting values we run a \textit{Markov Chain Monte Carlo} analysis for the 6 measurements of the averaged correlation function (see \cite{AB_voids2016} for the details). For each sub-mock we run 3 different \textit{Monte Carlo chains} to ensure the convergence of the best fitting values. For each mock, the likelihood is computed from scales $[1.5,45]\rm Mpc.h^{-1}$, along and across the line of sight in bins of $3\rm Mpc.h^{-1}$. 

\medskip
In Fig.~\ref{figBF_f}, dark green square,  we can see the best fitting values of the averaged 6 sub-mocks $f\sigma_8$ for the criteria of Eq. \ref{c1},\ref{c2},\ref{c3} where we use $\delta^j_{\rm T,G}(j,R=R_v)\geq \delta_{\rm ridge}$ for Eq.\ref{cp1} with $\delta_{\rm ridge}=0.1$. The error bars show the mean standard deviation across the 6 sub-mocks: $\sigma_6/\sqrt{6}$. The solid black line show the fiducial cosmology of the mocks ($f\sigma_8 \sim 0.376$). We see that on average the GSM recover the fiducial cosmology. This is consistent with the results presented in \cite{AB_voids2016} using galaxy mocks who have used galaxy mocks and fit for the linear bias. 

\medskip

In \cite{Hammaus2015JCAP}, the authors shown the same consistency of the growth rate only for certain void sizes. Depending on the void sizes, the void profiles show different features: the small voids have a high density on the ridge while the large voids have no ridge at all. In this work we can study the effect of the ridge at a fixed void radius, by keeping the void criteria of Eq. \ref{c1},\ref{c2},\ref{c3} and changing \ref{cp1}. For instance, changing Eq.\ref{cp1} to $\delta^j_{\rm T,G}(j,R=R_v)\geq \delta_{\rm ridge}$,  we can test the effect of the ridge by using $\delta_{\rm ridge}=-0.1,0,1,0.2,0.4,0.5,0.6$. 
The results are show in Fig.~\ref{figBF_f} by the blue, dark green, green, orange, pink and red squares respectively. Interestingly, the mean best fitting value for $f\sigma_8$ (averaged over the 6 sub-mocks), is consistent with the fiducial cosmology, at $1\sigma$ only for 2 cases ($\delta_{\rm ridge}=0.1,0.5$). However the scatter around the mean value is larger for $\delta_{\rm ridge}=0.5$. Hence, $\delta_{\rm ridge}=0.1$ gives the best constraint: $f\sigma_8=0.366\pm 0.04$. 
The other two extreme cases ($\delta_v=-0.1,0.6$) have a similar uncertainty but they also show a systematic error in the inferred value of $f\sigma_8$. This qualitatively highlights the importance of selecting voids when probing the growth rate, particularly the amplitude of the ridge. A more quantitative study would go beyond the scope of this work (see \cite{Chuang16} who also point some systematics error with the GSM in the context of redshift space distortion around voids). Overall, the criteria presented to identify voids can easily be used to different galaxy surveys (e.g. \cite{AB_voids2016}) and be tuned to study non-standard cosmologies (e.g. enhance departure from GR as we qualitatively study in sec.~\ref{Appli1}).

\section{Conclusion}\label{conclu}

In this work we present a new method to quickly test the imprint of non-standard cosmology using MCRW for a log-normal distribution. We focus on the departure from GR in the void density profiles for an $f(R)$ gravity model. In order to do so, we introduce flexible criteria to identify the random walks that mimic void profiles. 

\medskip

We find interesting results: our method can reproduce the qualitative features of the $f(R)$ gravity imprints in the void profiles found in \citep{ABPW}, without using a full N-body simulation. Furthermore, the departure from GR is sensitive to the type of voids, highlighting the importance of the ridge when identifying voids. 

\medskip

In addition, we test how flexible density criteria can be used to identify voids in real galaxy surveys. It would not be the case if the matter of our Universe would be randomly distributed. We also show how the flexibility to identify voids is important when probing the growth rate using redshift space distortions around voids.  

\medskip

Finally this work can be extended to different approaches, for instance it would be interesting to explore further the correspondence between the initial density criteria that lead to voids identified in the MCRW and N-body simulations. We could also test which are the initial density fluctuations that result in voids, for different cosmologies. This could potentially be used to model the abundance of voids. It would also be interesting to test how the voids identified with the criteria used in Sec.~\ref{sec2} change in a non-standard cosmology where the clustering properties of the matter are different (e.g.\ warm dark matter).
Our MCRW method can also be used to study the statistical properties of overdense fluctuations in the cosmic web such as peaks.

\medskip
The code used to find the voids in this work is available on demand if the reader is interested in using the method described in Sec.~\ref{sec2} and is not afraid of using Fortran. 

\section*{Acknowledgements}
We warmly thank Prof. Chris Blake and Prof. Tamara Davis for reading this paper and giving me useful comments. This work was conducted by the Australian Research Council Centre of Excellence for All-sky Astrophysics (CAASTRO), through project number CE110001020. We also acknowledge support from the DIM ACAV of the Region Ile-de-France.

\bibliography{biblio_voids}

\end{document}